\documentclass[sigconf]{acmart}

\setcopyright{none}
\renewcommand\footnotetextcopyrightpermission[1]{}
\AtBeginDocument{%
  }

\acmConference[]{}{}

\usepackage{enumitem}
\setlist{topsep=2pt,itemsep=1pt,parsep=0pt}

\usepackage[capitalise,noabbrev]{cleveref}
\usepackage{hyperref}
\usepackage[most]{tcolorbox}
\usepackage{xspace}
\usepackage{enumitem}
\usepackage{booktabs}
\usepackage{siunitx}
\usepackage{graphicx}

\newcommand{\systemname}{\textsc{Creo}}
\newcommand{\dg}[1]{\textbf{DG{#1}}\xspace}
\newcommand{\dgone}{\dg{1}}
\newcommand{\dgtwo}{\dg{2}}
\newcommand{\dgthree}{\dg{3}}
\newcommand{\stage}[1]{\textit{#1}}

\begin{document}

\title[Creo]
{Creo: From One-Shot Image Generation to Progressive, Co-Creative Ideation}


\author{Zoe De Simone}
\affiliation{%
  \institution{MIT CSAIL}
  \city{Cambridge}
  \state{MA}
  \country{USA}
}
\email{zoed@mit.edu}

\author{Angie Boggust}
\affiliation{%
  \institution{MIT CSAIL}
  \city{Cambridge}
  \state{MA}
  \country{USA}
}

\author{Fredo Durand}
\affiliation{%
  \institution{MIT CSAIL}
  \city{Cambridge}
  \state{MA}
  \country{USA}
}

\author{Ashia Wilson}
\affiliation{%
  \institution{MIT CSAIL}
  \city{Cambridge}
  \state{MA}
  \country{USA}
}

\author{Arvind Satyanarayan}
\affiliation{%
  \institution{MIT CSAIL}
  \city{Cambridge}
  \state{MA}
  \country{USA}
}

\renewcommand{\shortauthors}{De Simone et al.}

\begin{abstract}
 Text-to-image (T2I) systems enable rapid generation of high-fidelity imagery but are misaligned with how visual ideas develop.
T2I generate outputs which make implicit visual decisions on behalf of the user, often introduce fine-grained details which can anchor the user prematurely, 
limiting users’ ability to keep options open early on, and cause unintended changes during editing that are difficult to correct and reduce users’ sense of control.

To address these concerns, we present \systemname{}, a multi-stage 
T2I system that scaffolds image generation by progressing from rough sketches to high-resolution outputs, exposing intermediary abstractions where users can make incremental changes.

Sketch-like abstractions invite user editing and allow to keep design options open when ideas are still forming, due to their provisional nature.
Each stage in \systemname{} can be modified with manual changes and AI-assisted operations, enabling fine-grained and stepwise control through a locking mechanism that preserves prior decisions so subsequent edits affect only specified regions or attributes. Users remain in the loop, making and verifying decisions across stages, while the system applies diffs rather than regenerating full images, reducing drift as fidelity increases.
A comparative study with a one-shot baseline shows that participants felt stronger ownership over \systemname{} outputs, as they were able to trace their decisions in building up the image. Furthermore, embedding-based analysis indicates that \systemname{} outputs are less homogeneous than one-shot results. These findings suggest that multi-stage generation, combined with intermediate control and decision locking, is a key design principle for improving controllability, user agency and creativity, and output diversity in generative systems.
\end{abstract}

\ccsdesc[500]{Computing methodologies~Artificial intelligence}
\ccsdesc[500]{Human-centered computing~Interaction design}

\keywords{human-AI collaboration; generative AI; text-to-image models; creativity support}


\begin{teaserfigure}
  \includegraphics[width=\textwidth]{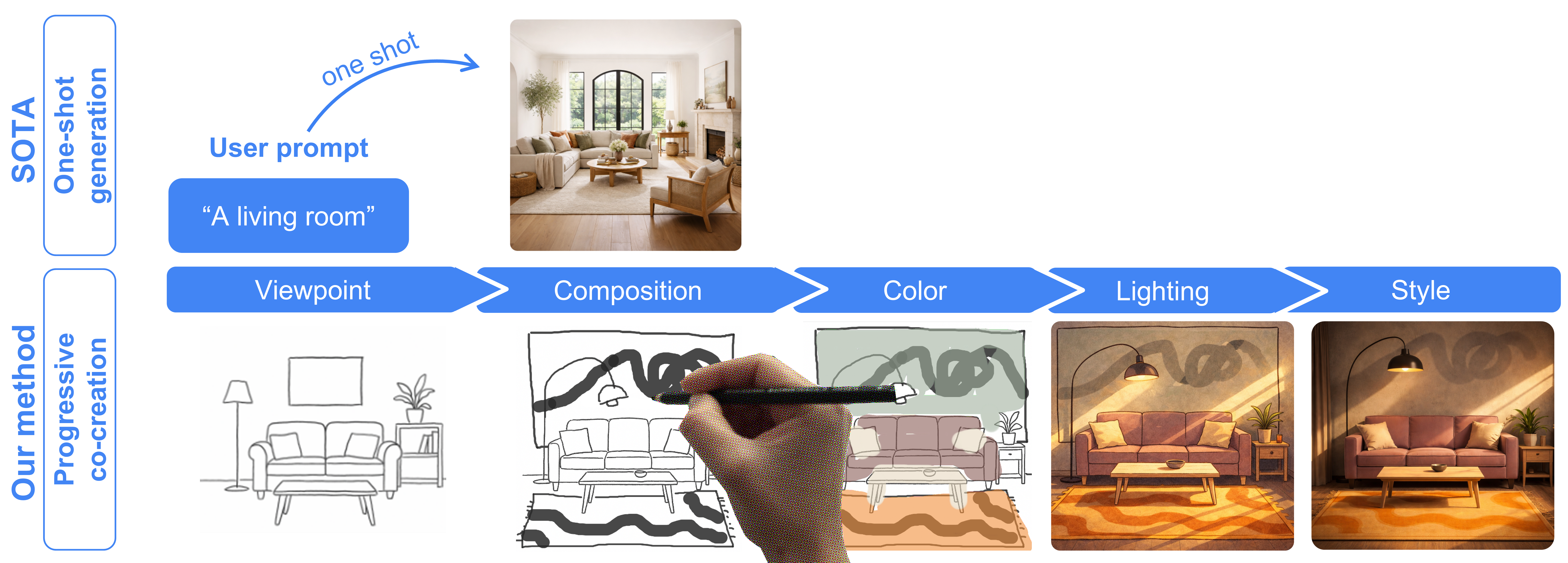}
  \caption{
  \systemname{} is a multi-stage image generation workflow. Unlike current text-to-image systems, \systemname{} starts from a rough sketch, allowing users to progressively make visual decisions, like viewpoint, composition, color, lighting, and style. 
  }
  \Description{}
  \label{fig:teaser}
\end{teaserfigure}


\maketitle

\section{Introduction}

Text-to-image (T2I) systems enable users to rapidly generate high-fidelity images from minimal prompts, lowering the barrier to image creation~\cite{ramesh2022hierarchical,saharia2022photorealistic,rombach2022high}. However, most systems \textbf{generate fully rendered images in a single step}, making implicit visual decisions about composition, lighting, and style before the user’s intent has fully formed.
These decisions can \textbf{prematurely anchor users} to details they have not yet considered and make edits feel like repairing model-made choices rather than developing one’s own ideas.

This interaction paradigm contrasts with how creative ideas typically develop. \textbf{Creative work rarely begins with a fully specified vision}~\cite{bigelow2023non}. Instead, creators \textbf{progressively refine} structure and appearance \textbf{over time}. Many visual workflows explicitly support this progression. For example, artists in animation and film begin with storyboards and rough pre-visualizations before committing to lighting and rendering~\cite{thomas1995illusion,okun2010vfx,parent2012computer}, while comics artists move from thumbnails to penciling, inking, and coloring~\cite{mccloud1994understanding,eisner2008comics,cohn2013visual}. These staged representations allow creators to focus on particular decisions while postponing others, enabling broad exploration without prematurely committing to fine-grained details. In contrast, single-shot \textbf{T2I systems collapse} these \textbf{decisions} into a single step, \textbf{forcing early commitment} and limiting exploratory control.

Together, this mismatch suggests a need for \textbf{generative systems that support progressive ideation and intent formation}, allowing users to develop and refine visual decisions over time rather than specifying outcomes upfront.

Prior work on controllable generation has expanded what aspects of an image users can specify—for example through structural guidance such as edges, poses, or layouts (e.g., ControlNet~\cite{zhang2023adding}). However, these approaches often still introduce highly detailed imagery early in the process. In contrast, we argue that \textbf{improving controllability requires structuring \emph{when} and \emph{how} visual decisions are introduced} during image creation.

To address this, we present \textbf{\systemname{}}, a T2I system that \textbf{scaffolds image generation} as a \textbf{multi-stage process}, progressing \textbf{from rough sketches to high-resolution outputs}. Rather than generating a fully rendered image immediately, \systemname{} decomposes image creation into a \textbf{sequence of stages} over which users retain control.

At each stage, users interact with \textbf{sketch-like intermediate representations} that are \textbf{intentionally} under-specified and \textbf{unfinished}, inviting user input and keeping design options open while ideas are still forming. As these representations evolve, they become progressively more detailed, allowing users to first establish high-level aspects such as viewpoint and composition, and then refine lower-level properties such as color, lighting, and style.

\systemname{} provides direct manipulation and AI-assisted tools that enable fine-grained control over specific aspects of the image. Users can \textbf{edit intermediate representations} by drawing, erasing, or masking regions and describing localized changes, allowing targeted modifications \textbf{without affecting unrelated elements}.

Through this staged interaction, users remain actively involved across the generation process, \textbf{making and verifying decisions} as the image progressively develops rather than reacting to fully specified outputs.
To support stable iteration, the system includes a \textbf{locking mechanism} that preserves prior decisions so that subsequent edits apply only to selected regions or attributes. Instead of regenerating entire images, \systemname{} applies incremental, diff-based updates, reducing unintended changes and drift as fidelity increases.

We evaluate \systemname{} through a \textbf{comparative study against a conventional one-shot generation} workflow. Participants reported a stronger sense of \textbf{ownership} over images created with \systemname{}, as they were able to \textbf{trace their decisions} in building up the image.
Participants engaged with the full range of stages and both direct and AI-assisted tools, despite this not being required, indicating that intermediate stages were not superfluous but provided distinct and necessary support for different aspects of the workflow. Additionally, embedding-based analysis shows that outputs from \systemname{} are \textbf{less homogeneous} than those produced by the baseline, suggesting broader exploration of the design space.
While baseline systems were often rated higher in immediate visual quality, \systemname{} \textbf{improved users’ sense of control, authorship, and ability to iteratively refine ideas}. Together, these findings suggest that structuring generation as a multi-stage process—combined with editable intermediate representations and mechanisms for preserving decisions—is a key design principle for improving controllability, user agency, and output diversity in generative systems.

More broadly, this work highlights the importance of \textbf{supporting progressive intent formation}, where systems help users develop ideas over time rather than requiring them to specify outcomes upfront. By aligning generative workflows with the iterative nature of creative thinking, such systems can mitigate well-known behavioral biases such as anchoring effects and premature commitment
~\cite{green1989cognitive,jansson1991design,wadinambiarachchi2024effects,doshi2024generative}, enabling more exploratory and user-driven ideation. 

\section{Background and Design Goals}
\label{sec:relatedwork}

Prior research across cognitive science, creativity support tools, and generative image systems points to a \emph{mismatch between how people form visual ideas and how current T2I workflows generate images}. 
In creative practice, ideas often develop gradually through intermediate representations that help creators refine one aspect of an image at a time. For example, artists will isolate and test different color palettes on small regions of an image prior to choosing a direction for the entire image. In contrast, many T2I systems produce fully rendered outputs early, collapse multiple visual decisions into a single generation step, and rely heavily on prompt-based interaction to control visual decisions. 

\subsection{Human Requirements of Visual Ideation}

\textbf{Early visual ideation is intentionally underspecified.}
In creative practices, early visual ideas are often deliberately underspecified. Cognitive science research shows that people imagine scenes without committing to attributes such as color, texture, or background until later in the process~\cite{bigelow2023non}. This ambiguity is beneficial because it allows multiple candidate interpretations of a scene to remain viable until later design decisions constrain them and helps preserve flexibility while ideas are still forming.

Sketching makes this process visible. Drawing externalizes evolving thought~\cite{fan2023drawing}, and early sketches are treated as exploratory~\cite{purcell1998drawings}. Classic accounts describe sketching as a reflective process of seeing, interpreting, and revising~\cite{tversky2002sketches,goldschmidt2014modeling,schon1983reflective}. 
Line drawings preserve structure while suppressing detail, making them well suited for reasoning about form without committing to appearance~\cite{hertzmann2020line,livingstone2002vision}. 

Together, these findings suggest that ideation benefits from representations that preserve ambiguity.

\medskip
\noindent
\textbf{Intermediate representations support dimension-specific reasoning.}
Creative workflows typically progress through representations that isolate specific types of decisions. For example, artists establish composition before refining lighting or color~\cite{dow1914composition,loomis1947creative}, and workflows in animation, film, and comics similarly move from structure to appearance~\cite{thomas1995illusion,okun2010vfx,parent2012computer,mccloud1994understanding,eisner2008comics,cohn2013visual}. This ordering reflects dependencies between decisions, as later properties rely on a stable structural foundation~\cite{purcell1998drawings,verstijnen1998sketching,brun2016designing}.

Intermediate representations also support attention and reasoning by limiting what is being decided at a given time~\cite{goodwin961994}. Creativity support tools have long leveraged similar strategies, enabling users to begin with coarse sketches or low-fidelity wire frames and progressively refine them~\cite{igarashi1999_teddy,bae2008ilovesketch,swearngin2018scout,10.1145/2702123.2702149,iarussi2013drawing,fernquist2011sketch}. 
Recent systems such as DrawMyPhoto~\cite{williford2019drawmyphoto} and ArtKrit~\cite{ma2025artkrit} externalize dimensions such as composition, value, and color to support focused reasoning. Drawing assistance tools guide contour structure or proportion without requiring fully rendered outputs~\cite{iarussi2013drawing,fernquist2011sketch}. Other tools provide abstractions for working with color and tone, including histogram views~\cite{chevalier2012histomages} and direct-manipulation palette systems~\cite{shugrina2019color,shugrina2017playful}.

Across these systems, intermediate representations act as cognitive scaffolds that support incremental, focused reasoning. This work suggests that visual ideation is inherently staged: creators progressively and independently refine image dimensions.

\subsection{Challenges in T2I Model Workflows}
\label{sec:related-model}

\textbf{Early specificity shifts ideation toward reactive editing and constrains exploration.}
Most T2I systems generate fully specified, high-fidelity images from the outset. Even minimal prompts can produce detailed commitments to lighting, texture, background, and style. While visually compelling, these outputs introduce many decisions that users may not have intended or even considered~\citep{hopkins2025chatbot}, often before their intent has stabilized.

As a result, interaction becomes reactive. Users typically iterate by modifying prompts and regenerating full images, but each iteration still presents a complete, highly detailed output. Rather than developing ideas through intermediate representations, users adapt their thinking in response to model-produced artifacts. Early exposure to polished outputs can also create a false sense of completion, where users accept compelling images despite misalignment with higher-level goals such as composition or viewpoint.

Correcting these discrepancies often requires localized edits like masking, inpainting, or prompt adjustments. This shifts the creative process away from concept formation toward repairing model-generated artifacts. Prior work describes this dynamic as a “gulf of envisioning,” where users must specify details they cannot yet imagine and interpret unpredictable changes in generated outputs~\cite{agrawala2023unpredictable,subramonyam2024bridging}.

Early specificity also shapes exploration through well-established cognitive mechanisms. Design fixation shows that exposure to example solutions can anchor designers on particular features, even when those features are incidental~\cite{jansson1991design}. Generative model outputs can play a similar role: users may begin ideation by reacting to generated images rather than constructing ideas incrementally. Recent work suggests that this can reduce divergent thinking~\cite{wadinambiarachchi2024effects} and lead to more homogeneous outputs across users~\cite{doshi2024generative}.

Current interaction patterns amplify these effects. Interfaces often present grids of polished outputs or support prompt refinement~\cite{feng2023promptmagician,brade2023promptify}, encouraging comparison between finished artifacts rather than early exploration. Hybrid systems that combine sketches with high-fidelity generation may similarly introduce detailed imagery before structural decisions are fully developed~\cite{zhang2023adding,sarukkai2024block}.

Together, these findings reveal a mismatch between the timing of model-generated detail and the development of user intent. When highly specific visual commitments are introduced too early, they can anchor exploration, shift interaction toward reactive editing, and reduce diversity of outcomes. This suggests that generative systems should align the level of visual specificity they produce with the user’s stage of ideation.



\medskip
\noindent
\textbf{Visual decisions are entangled in current interaction paradigms.}
In most T2I systems, decisions such as viewpoint, composition, color, lighting, and style are resolved simultaneously within a single generation step. Because these attributes are entangled, modifying one aspect often unintentionally affects others. For example, changing lighting conditions may alter color, mood, or object appearance, requiring repeated regeneration. 

Prompt-based interaction further complicates this process. Users must translate evolving visual ideas into explicit language, even when intent is still tacit~\cite{schon1983reflective,tversky2002sketches}. Recent work on controllable generation expands the range of attributes users can specify, for example through structural guidance or localized editing~\cite{zhang2023adding,sarukkai2024block,voynov2023sketch,tang2025_sketch_segmentation,lee2025_3d_sketching_car,alzayer2025magic,zhang2025_digital_painting,shi2026_notational_animating}. However, these approaches primarily address \emph{what} can be controlled, while less often addressing \emph{when} different visual decisions should be introduced.

\subsection{Design Goals}

These observations point to a consistent mismatch: human ideation proceeds through staged, partially specified representations, while current T2I systems introduce detailed, entangled decisions early. From this mismatch we distill three design goals (\dgone–-\dgthree), which inform the design of \systemname{} described in Section \ref{sec:system}.

\begin{enumerate}[label=\textbf{DG\arabic*:}, wide=0pt, itemsep=0.4em]
    \item \textbf{Introduce visual detail progressively.} Generative systems should align the level of detail they produce with the user’s stage of ideation, enabling early exploration before committing to fine-grained appearance, and without prematurely committing a user to details they have not yet reasoned about.

    \item \textbf{Decompose image creation into separable decisions.} Inspired by artistic practices, generative systems should structure image creation around distinct visual dimensions (e.g., composition, lighting). This allows users to incrementally build up their image as intent forms, refining one aspect without unintentionally altering others. Such separation enables localized reasoning and editing while preserving previously made decisions.

    \item \textbf{Support interaction through representations accessible to both humans and models that invite editing.} Generative interfaces should employ intermediate visual representations that are both intuitive for users to manipulate and meaningful for models to condition on. Rather than requiring users to edit fully rendered images or specify pixel-level changes through prompts, systems should expose abstractions that allow users to control one aspect of the image at a time and are easily accessible and editable.
\end{enumerate}

\section{\systemname}
\label{sec:system}

\begin{figure*}[t]
\centering
\includegraphics[width=\linewidth]{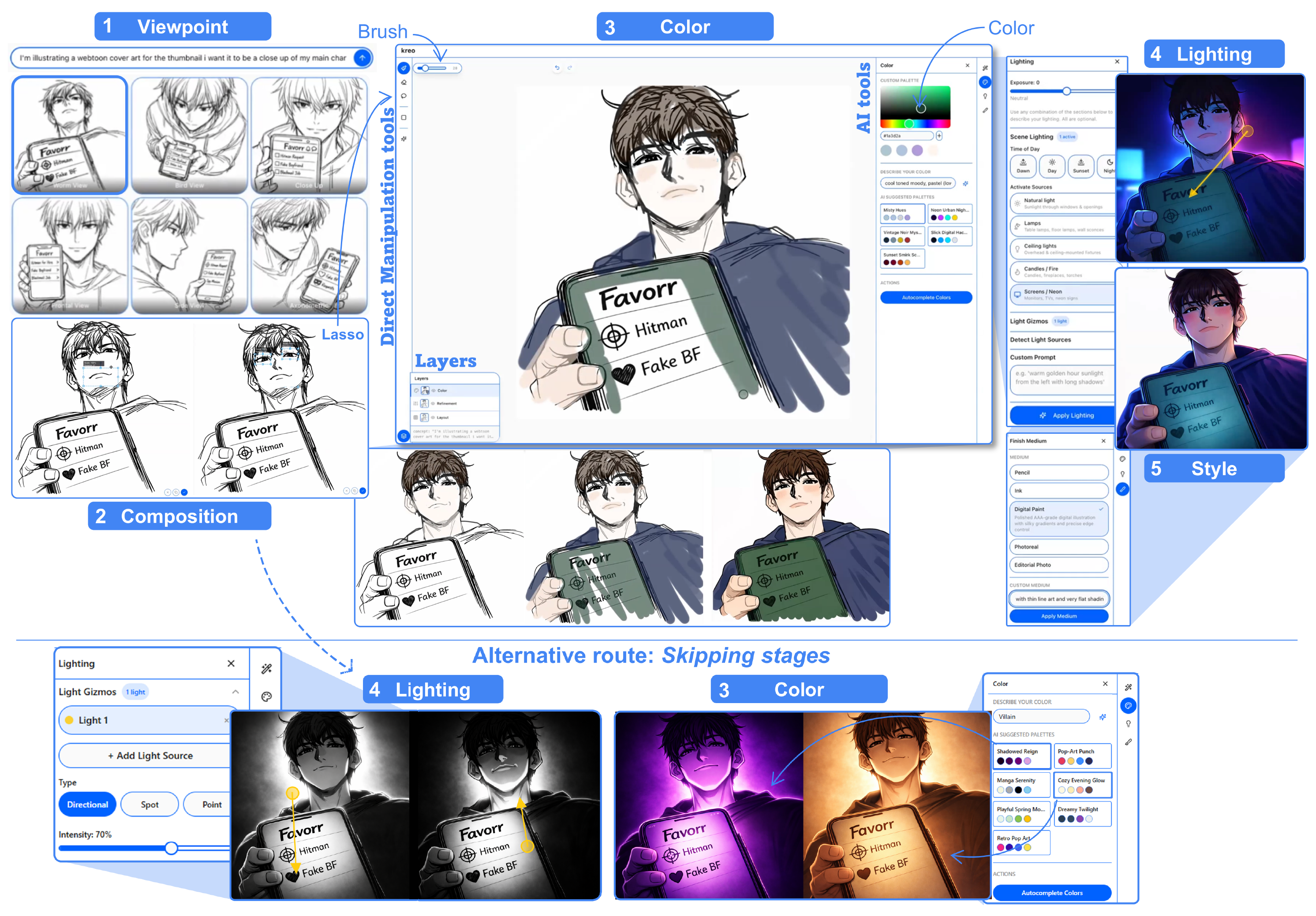}
\caption{
\systemname{} decomposes image generation into multiple stages. From a prompt, it generates (1) multiple \stage{viewpoints}, after which the illustrator (2) refines \stage{composition}, (3) \stage{color}, (4) \stage{lighting}, and (5) \stage{style} in any order. Stages can be completed in any order.}
\label{fig:comics_figure}
\end{figure*}

To bridge the disconnect between human visual ideation and current image generation workflows, we introduce \systemname{} (\cref{fig:comics_figure}).
\systemname{} is a text-to-image system that supports progressive ideation (\dgone) by structuring image generation as a sequence of creative decisions (\dgtwo).
Rather than generating a fully specified image in a single step, \systemname{} begins with sketches that users progressively refine into high-fidelity images (\dgthree).

\subsection{Multi-Stage Image Generation}
\label{sec:multistage}

We designed \systemname{} to give users fine-grained control over image generation without requiring them to specify every visual property at the outset.
To do so, \systemname{} decomposes text-to-image generation into five stages: \stage{viewpoint}, \stage{composition}, \stage{color}, \stage{lighting}, and \stage{style}.
In each stage, users make the design decisions that are relevant to that part of the creative process, such as sketching new objects during \stage{composition} or choosing a \stage{color} palette.
This design facilitates progressive ideation (\dgtwo), allowing users to build up the final image in a way that mirrors the creative process, starting with low-level structural decisions and then refining higher-level aesthetic details.

Importantly, \systemname{}'s stages are independent, allowing users to explore one aspect of their image's design at a time.
While standard text-to-image systems always output a fully rendered image (e.g., adding color even when none is specified), \systemname{} better matches the level of detailed described in the prompt. 
For instance, if a user has not entered the \stage{color} stage, then the image will remain in grayscale while they make decisions about \stage{lighting},  keeping options like color and style open for later refinement.

\systemname{}’s stages are also additive, enabling users to progressively build a high-fidelity image.
As users move through the stages, their decisions compound, and the image evolves from a black-and-white sketch to a fully rendered scene.
Users work with sketches in \stage{viewpoint} and \stage{composition}, add hues in \stage{color}, define depth in \stage{lighting}, and develop an aesthetic in \stage{style}.
As a result, \systemname{} allows users to work through their creative process, yielding an image that fully results from a user's intent rather than an AI model's assumptions.

By structuring image creation into stages, \systemname{} isolates user edits.
A central challenge in text-to-image editing is that modifying one aspect of an image often unintentionally affects others.
Instead of modifying a fully rendered image and resolving unintended changes, users verify their \stage{composition}, \stage{color}, and \stage{lighting} choices step by step.
This staged process helps prevent errors from compounding as the image becomes more detailed.

By decomposing image generation into stages, \systemname{} aligns with established creative processes.
The five stages are inspired by documented workflows in illustration, studio art, and architectural rendering~\citep{SzucIllustrationProcess, ArtistsIllustratorsChildrensBook, Nicolaides1941NaturalWay, Ching2014ArchitecturalGraphics}, where, practitioners often establish composition before committing to color, lighting, and style.
While individual creative processes vary based on the creator and their task, we distill these recurring practices into a compact set of five stages.
As a result, \systemname{}'s stages act as a minimal grammar of visual ideation, allowing users to control their creative process while leveraging the efficiency of image generation models.


\subsection{Staged Representations and Tools}
\label{sec:sketch}

To support creative processes, \systemname{} defines stages that allow users to progressively introduce visual details, like \stage{lighting} and \stage{color}.
However, progressive ideation is not only about matching the visual detail in the image to the user's intent, but allowing users to meaningfully edit these underspecified images (\dgone).

To do so, \systemname{} maps each stage (\cref{sec:multistage}) to an intermediate representation.
Intermediate representations reflect the types of visual design decisions a user makes in that stage.
Instead of editing a rendered image, users work on representations that isolate specific aspects of the image and remove dependencies on others.
For instance, the \stage{composition} stage operates over black-and-white sketches, while \stage{lighting} adds shading to the sketch, and only \stage{style} renders a high-fidelity image.
This allows users to reason directly about the decision at hand without being constrained by details that they have not yet specified.

In particular, \systemname{} uses sketch-based representations to capture structure without committing to appearance.
Sketches reduce unrequested detail not by hiding it, but by not introducing it in the first place.
This matches users' intent by removing dependencies on details that have not yet been decided, and supports progressive ideation through changes in representation rather than incremental refinement of a single image.
For example, in a sketch, repositioning an object involves only adjusting simple shapes, rather than reconciling perspective, shadows, or occlusion.
Moreover, since sketches are perceived as provisional they encourage users to modify and explore the representation rather than treat the image as final.

Each stage and its representation (e.g., black-and-white sketch) is mapped to a set of interaction techniques that operate on that representation (e.g., lasso and move).
This pairing allows users to act directly on the aspect of the image they are currently refining (\dgthree).
While we initially explored a unified set of tools across all stages, this made interaction cumbersome as tools designed for detailed images (e.g., lighting) are difficult to apply to simplified representations and sketch-based tools do not translate well to appearance-level editing (e.g. what is the effect of brushing on lighting or how should drawing alter a high-resolution image).

Instead, each stage provides tools that match its representation. 
For instance, in a sketch-based stage like \textit{composition}, users work with tools such as drawing, erasing, and transforming shapes to define structure.
As the image becomes more detailed, \systemname{} introduces tools for applying color through brushes or regions, and for adjusting lighting through directional and intensity controls.
By scoping tools to each stage, users can act directly on the aspect of the image they are currently refining, without needing to manage unrelated details.
This reduces interference between decisions and keeps interaction aligned with the current level of abstraction.

Within each stage, \systemname{} combines direct manipulation and AI-assisted tools.
We combine these two types of interactions to balance between giving users precise editing control and allowing them to delegate tedious edits to an AI model.
As a result, users can provide partial input, such as rough sketches, color strokes, or masks, and rely on the system to complete or refine the result. 
For example, a user may block in approximate color regions and allow the system to propagate them into object boundaries or specify a lighting direction and let the model infer shading.
This allows users to control the level of effort they invest, focusing on decisions they care about while delegating others to an AI model.

To limit unintended changes, \systemname{} supports localized edits through masking. 
Users can select specific regions of the image and apply changes only within those areas.
Model-generated updates are composited back into the existing image, leaving the rest unchanged.
This allows users to refine individual elements—such as adjusting the color of an object without affecting the surrounding scene.

\begin{figure}[t]
\centering
\includegraphics[width=\linewidth]{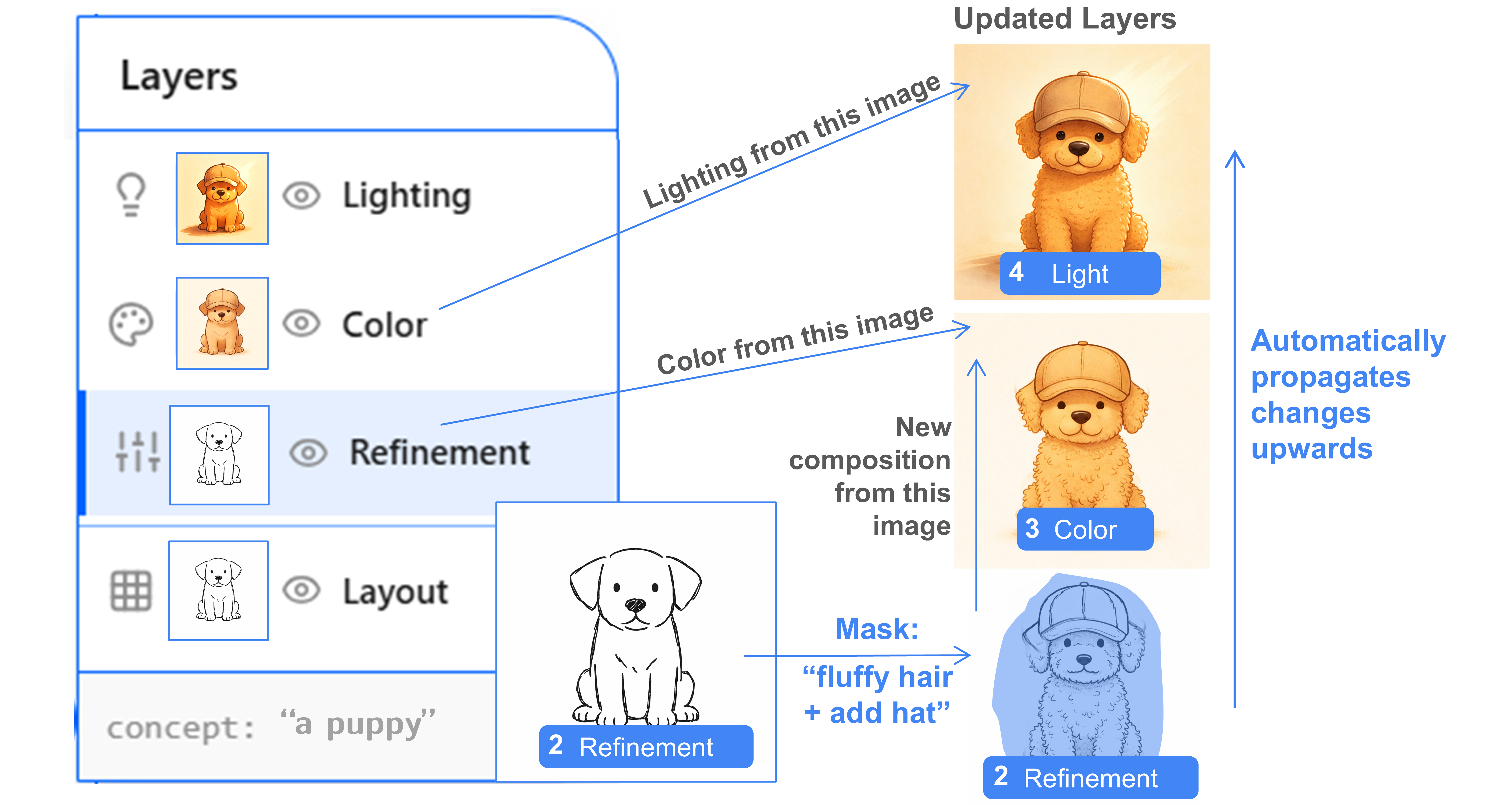}
\caption{
\systemname{} supports non-linear creative workflows by combining edits in earlier stages (e.g., adding a hat to the \stage{composition}) with existing upstream decisions (e.g., dog \stage{color}).
}
\label{fig:kreo_propagation}
\end{figure}

\subsection{Interaction Workflows}

Although \systemname{} is organized into stages, creative workflows are not strictly linear. 
Users often revisit earlier decisions after seeing later results, requiring the system to support revision without restarting.

To support non-linear workflows, \systemname{} provides a persistent preview of the final image alongside the decisions in each stage (\cref{fig:kreo_propagation}).
Users can move between stages in any order and revise earlier decisions (Figure~\ref{fig:decision_propagation_diagram}), such as coloring an image and then returning to composition to add a new object or change the layout.
This keeps different stages conceptually separate, while always revealing the current composed result.
In informal pilot studies, we found that using \systemname{}'s set of unordered stages, as opposed to a timeline of sequential edits~\citep{willis2021fusion}, helped users understand the impact of their edits without overloading the workspace.

To ensure users can revisit stages without losing their existing progress, \systemname{} implements \textit{re-propagation}.
When a user edits an earlier stage, those changes are propagated forward to downstream stages while preserving existing decisions where possible.
For example, in \cref{fig:kreo_propagation}), when a user modifies a dog's fur type and adds accessories after specifying \stage{color} and \stage{lighting}, they will see the updated \stage{composition} rendered with the previously defined palette and illumination.
By scoping changes to a specific aspect of the image and referencing previous intermediary representations as locked decisions, the system reduces unintended interactions between decisions and maintains consistency across stages.

Re-propagation is implemented by ensuring that every generation step is conditioned on both the current stage as well as a the user's decision state.
The decision state includes the current composed image, decisions in earlier stages, and stage-specific instructions defining which attributes are editable (\cref{fig:decision_propagation_diagram}).
Rather than regenerating images from scratch, \systemname{} incrementally updates the current image while locking the user's previous decisions.
This allows users to compare alternatives, revert edits, and branch from earlier decisions without losing progress.

\subsection{Design Rationale and Tradeoffs}

Designing \systemname{} required navigating several tradeoffs.
Below we describe the key tensions that shaped our design.

\subsubsection{Progressive Detail vs. Editable Representations.} A central tension is how to progressively introduce visual detail (\dgone) while ensuring users can meaningfully manipulate incomplete images (\dgthree). 
A natural approach is to vary the level of detail within a single image, for example generating a high-resolution output while simplifying underspecified regions.
For example, given the prompt ``\textit{a living room with an orange sofa near a window},'' where the sofa is well specified but the surrounding layout remains ambiguous, a system might fully render the sofa while blurring the background.

In practice, however, this does not support progressive ideation.
Even when regions are blurred, users must still interpret and edit a fully rendered scene.
For example, repositioning the sofa requires reasoning about perspective, shadows, and surrounding objects even in visually de-emphasized areas.
As a result, users often avoid direct manipulation of fully rendered images and instead regenerate outputs or adjust prompts, reinforcing a reactive workflow.

This contrast is clear when compared to \systemname{}'s sketch-based interaction.
Editing a sketch of the same scene decreases the cognitive and interaction cost. 
Repositioning the sofa only requires adjusting simple outlines without needing to account for lighting or texture.
Moreover, while rendered images appear ``finished'' and discourage editing, sketches appear provisional and invite modification.

This reveals a limitation of applying \dgone in isolation.
Progressively revealing detail is insufficient if interaction remains tied to pixel-level representations that are challenging to edit.
As long as users operate on rendered images, they must reason about appearance rather than structure.


\subsubsection{Speed vs. Structured Control.}
Another tension lies between rapid iteration and decomposed control.
One-shot generation enables fast exploration by producing complete images from a single prompt, but entangles multiple decisions and makes targeted edits difficult.
Decomposing generation into stages introduces structure and control, but increases interaction overhead.

This raised the question of how many stages to include and what decisions each should expose.
We initially explored designs that mapped sketches directly to fully rendered outputs, similarly to existing systems such as ControlNet~\citep{zhang2023adding} and Block and Detail~\cite{sarukkai2024block}.
While this provided control over the composition of the image, it left many other aspects of the image such as color, lighting, and style up to the model, and we found it anchored details prematurely and limited control over the aesthetic aspects of the image.

Instead, we grounded \systemname{}'s stage decomposition in studio art practices (e.g., thumbnailing, refinement, color studies, shading)~\cite{loomis1947creative,mccloud1994understanding,eisner2008comics,dow1914composition} and extended this progression with a final rendering stage.
Informal feedback indicated that users consistently valued control over composition, color, and lighting, even if their priorities differed
We therefore adopted this set of stages as a balance between expressive control and interaction complexity.

\subsubsection{Linear vs. Non-Linear Workflows.}
Creative workflows are non-linear, and users frequently revisit earlier decisions after observing downstream results.
Thus, while staged decomposition introduces clarity, forcing a sequence of stages would be misaligned with practice. 
To support non-linear workflows, we explored a timeline-based approach~\citep{willis2021fusion}, but found that organizing interaction temporally made it difficult to distinguish between types of visual decisions and how edits would affect the image.
Instead, \systemname{} adopts a semantic organization~\cite{photoshop2026} where each stage corresponds to a visual decision (e.g., \stage{color}) but users can move between stages in any order.


However, stage-based design introduces a cognitive challenge: revisiting earlier stages can feel like ``going back in time,'' requiring users to recall downstream decisions.
We explored showing multiple stages simultaneously, but found this visually cluttered. 
Instead, we provide a persistent preview of the final image, allowing users to understand the impact of edits without overloading the workspace.




\subsubsection{User Control vs. Automation.}
A final tension concerned balancing how much manual control to give users via direct manipulation tools vs. how to leverage AI model's abilities to alleviate busy work.
We address this by supporting multiple levels of control within each stage. 
Users can directly manipulate representations (e.g., drawing structure, applying color, specifying lighting direction) or provide partial input (e.g., masks, palettes, or high-level prompts) and rely on the model to complete the result.
For example, users may sketch approximate color regions and use AI-assisted tools to refine them, or specify a lighting “vibe” and allow the system to infer detailed illumination.
This flexibility gives users agency to allocate their effort, focusing on decisions they care about while delegating others—and accommodates different creative styles.


\subsubsection{General Purpose vs. Sketch-Specific Models.}
\systemname{} relies on T2I models to translate user edits into an updated image.
Since many stages rely on sketch-based representations, we initially explored using specialized sketch-generation models~\citep{Vinker_2025_CVPR}.
However, we found that prompt-based adaptations of general-purpose models were sufficient to produce sketch-like outputs, allowing us to use a single model family across all \systemname{} stages.

\section{User Study: One-Shot vs Progressive Generation Evaluation}
\label{sec:userstudy}

To evaluate whether structuring image generation into progressive stages influences creative ideation, we conducted a controlled user study comparing \systemname{}’s staged workflow with a conventional one-shot text-to-image interface. Our goal is to understand how these workflows shape (1) breadth of idea exploration, (2) users’ ability to make targeted changes without disrupting prior decisions, (3) when users commit to a design direction, and (4) their sense of authorship and whether they feel they are actively shaping the image or reacting to model-generated outputs.

\subsection{Study Overview}

We employed a within-subjects, counterbalanced design with two conditions:

\textbf{Progressive Staging (\systemname{}).} Participants used \systemname{}’s staged workflow, which decomposes image generation into sequential decision stages including viewpoint, composition, color, lighting, and rendering style. Participants could move freely between stages and revisit earlier decisions.

\textbf{One-Shot Baseline.} Participants used a multi-turn prompt-centric text-to-image interface (ChatGPT) that supported iterative prompt refinement, regeneration, and masked edits.

To reduce concept carryover between conditions and ensure participants were starting each task fresh conceptually, participants completed two related design prompts: \emph{design your ideal living room} and \emph{design your ideal kitchen}. Each participant completed both prompts, using a different tool for each prompt. Tool order and prompt order were counterbalanced across participants.

\subsubsection{Participants:}
We recruited 15 participants through posts in online (Reddit) and university communities focused on illustration, art, and design. Participants had prior experience in creative workflows across domains including graphic design and marketing (n=6), interior design and architecture (n=5), illustration and concept art (n=2), animation (n=1), and fine arts (n=1). Participants represented a mix of early-career practitioners (1–3 years, n = 4), to mid-level experience (4–7 years, n = 6), to highly experienced practitioners (8+ years, n = 5).
Participants had a range of prior exposure to generative AI image tools: 4 reported using them regularly, 6 occasionally, and 5 had limited or no active use. Sessions lasted approximately 60 minutes and were conducted remotely via video conferencing with screen sharing enabled. Participants received a gift card as compensation for their time.

\subsubsection{Study Protocol}
Sessions began with a brief discussion of participants’ creative workflows, use of T2I generative AI tools in their work. Participants then completed both conditions in counterbalanced order. Before each condition, the moderator introduced the interface. Participants were asked to think aloud while working for approximately 14 minutes per condition. After each condition, participants reflected on their experience, followed by a final composition discussion between the two workflows. Additional details on materials, task instructions, counterbalancing, and moderation are provided in Appendix~\ref{sec:study_appendix}.

\subsection{Analysis Overview}

We conducted a mixed-methods analysis combining structured interaction logs with qualitative analysis of participants’ verbal reflections. Our analysis is organized around three research questions corresponding to exploration, control, and workflow structure.

From screen recordings, we reconstructed interaction sequences and segmented them into discrete actions (e.g., \textit{construct}, \textit{evaluate}, \textit{generate}, \textit{repair}). Each action was annotated with its intent (\textit{on-intent}, \textit{pivot}, \textit{drift}) and whether it was \textit{user-driven} or \textit{model-led}, and whether it constituted a \textit{direction change} (a major shift in the overall concept or design trajectory). Actions were grouped into \textit{iterations} to capture branching exploration. 

These annotation categories were derived iteratively from our research questions, aiming to operationalize constructs such as exploration, control, and workflow structure. We refined the schema through multiple passes over a subset of sessions, adjusting definitions to ensure they were both interpretable from observable behavior and consistently applicable across participants. To scale annotation, we implemented an LLM-based coding pipeline that labeled each action a user took according to this schema. We validated this process by manually cross-checking a subset of annotated sessions against the original recordings, ensuring consistency between automated labels and observed user behavior.

Each annotation is used to derive session-level metrics (e.g., direction changes, drift, repair actions), which we define briefly below and describe in full in Appendix~\ref{sec:study_appendix}.

\textbf{RQ1: Exploration and anchoring.}
To characterize ideation breadth and commitment, we measure the number of \textit{direction changes} per session, individually coded from the study recordings, and the number of distinct \textit{iterations} (i.e., separate exploration branches). We additionally compute the proportion of actions labeled as \textit{drift} (unintended deviations introduced by the model) and \textit{pivot} (intentional redirection by the user). 

As a proxy for anchoring, we compute the L2 distance between unit-norm CLIP embeddings of the first and final images in each session. This summarizes how much endpoint appearance (in CLIP space) drifts from the initial frame: larger distance means the last image lies farther from the first in the model’s representation; smaller distance means the session remains closer to its starting point, which we treat as stronger anchoring in embedding space.

\textbf{RQ2: Control and predictability.}
To examine interaction control, we measure the proportion of \textit{construct} actions (direct content specification or modification) and \textit{evaluate} actions (inspection of outputs). We approximate perceived agency as the proportion of \textit{user-driven} actions (actions where users explicitly specify or modify content). We quantify \textit{revision burden} as the ratio of \textit{repair} actions (corrections of unintended changes), and measure unintended side effects via the rate of \textit{invariant violations}, defined as edits that unintentionally alter aspects outside their intended scope (e.g., changing layout during a color edit).

\textbf{RQ3: Workflow structure.}
To understand how participants appropriate staged interaction, we analyze iteration patterns and stage transitions in \systemname{}, including \textit{revisiting} earlier stages and \textit{skipping} intermediate stages. We also summarize stage usage as the proportion of actions occurring within each stage and the proportion of participants who engaged with each stage.

\textbf{Qualitative analysis.}
We analyzed think-aloud protocols and post-task interviews using thematic coding to capture participants’ experiences of authorship, control, predictability, and commitment. These qualitative themes are used to interpret the behavioral patterns observed in each condition, linking quantitative measures (e.g., constructive vs.\ evaluation-heavy behavior, direction changes, repair actions) to participants’ reported experiences of constructing, steering, or reacting to model outputs.

All metrics are computed at the session level and aggregated across participants, and are based on action counts and proportions rather than time-based metrics. Full annotation and coding details are provided in Appendix~\ref{sec:study_appendix}.

\subsection{User Study Results}
\label{sec:userstudy_results}

Our mixed-methods analysis showed that \systemname{} and the one-shot baseline (GPT) supported different modes of ideation. GPT positions users as reacting to finished outputs, while \systemname{} enabled users to progressively construct images through editable intermediate states, increasing perceived agency and authorship, but also introducing additional revision work due to its persistent structure. One-shot baseline outputs were regarded as highly-polished, and unintended changes were perceived as aesthetic recommendations, while \systemname{} outputs were perceived as provisional and editable. Below, we present four findings showing how staged interaction changes both what users can control and how they relate to the artifact they are creating.

\begin{table}[t]
\centering
\sisetup{separate-uncertainty, table-format=2.1(2.1)} 
\caption{\systemname{}'s staged generation outperforms GPT's one-shot prompting across exploration and control metrics ($\mu \pm \sigma$).}
\label{tab:metrics}
\resizebox{\linewidth}{!}{%
\begin{tabular}{@{} l S S @{}}
\toprule
\textbf{Metric} & {\textbf{\systemname{}}} & {\textbf{GPT}} \\
\midrule
\textit{Exploration \& Anchoring} \\
Direction changes (\# changes per session) & 1.6(10) & 0.4(5) \\
Exploration breadth (\# iterations per branch) & 3.3(18) & 2.5(17) \\
Concept drift (\% unintended drift actions) & 14.4(102) & 14.2(89) \\
Intentional pivots (\% pivot actions) & 3.9(34) & 2.9(43) \\
\addlinespace
\textit{Control \& Predictability} \\
Constructive engagement (\% construct actions) & 17.7(80) & 4.3(76) \\
Evaluation-heavy behavior (\% evaluate actions) & 35.7(58) & 39.4(106) \\
Behavioral agency (\% user-driven actions) & 55.0(118) & 25.8(125) \\
Revision burden (ratio of repairs to total actions) & 0.1(1) & 0.0(1) \\
Unintended changes (\% invariant violations) & 10.9(103) & 0.0(0) \\
\bottomrule
\end{tabular}
}
\end{table}

\subsubsection{Provisional representations reduce anchoring and support redirection}
\begin{figure}[t]
\centering
\includegraphics[width=\linewidth]{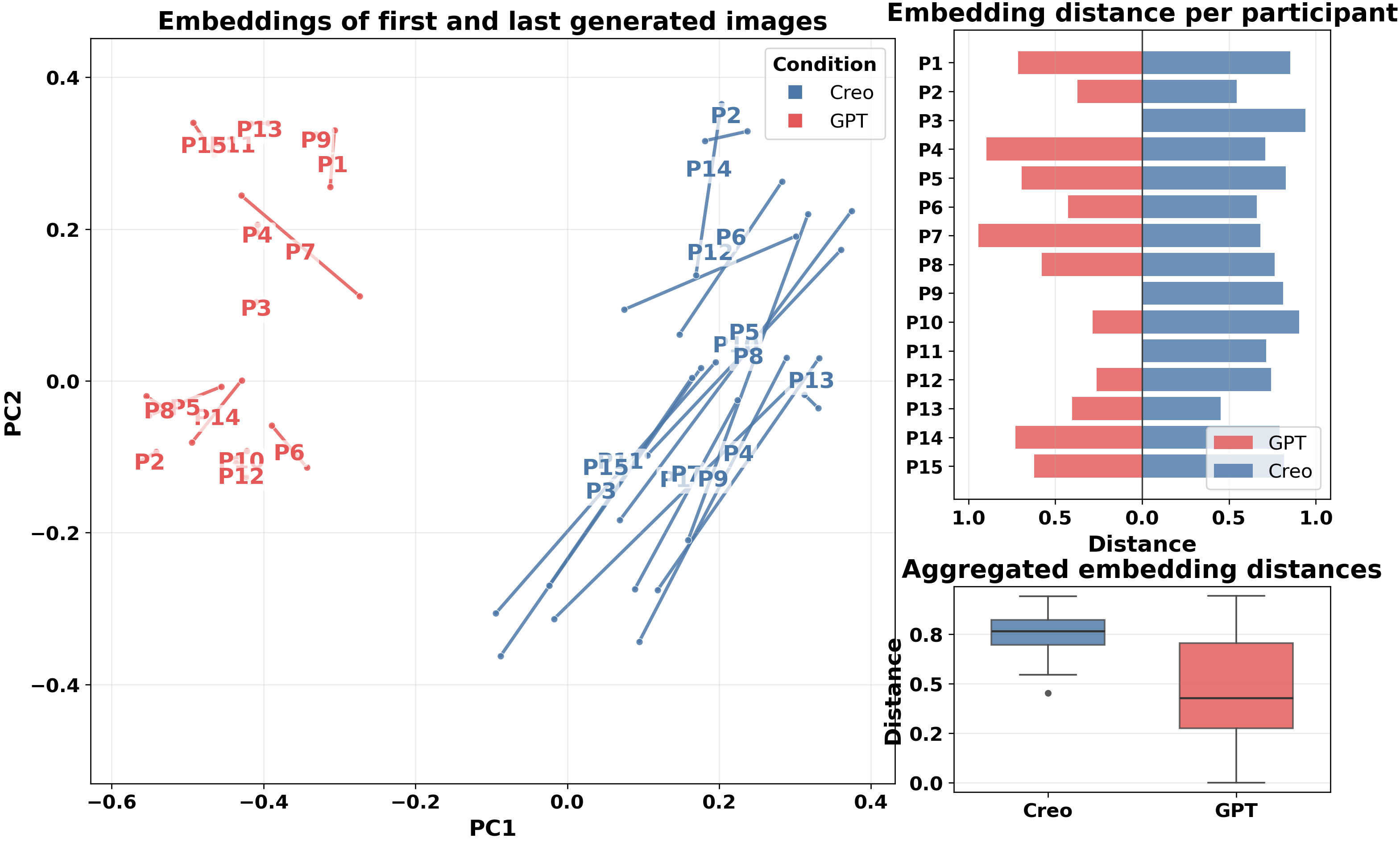}
\caption{
\systemname{} reduces design anchoring and homogenization. Visual embeddings of the images that users produced using \systemname{} are less tightly clustered than those from GPT.
}
\label{fig:kreo_results_homogenization}
\end{figure}

Participants were less anchored to early outputs in \systemname{} than in GPT, both behaviorally and in how they described the workflow. \systemname{} sessions included more \textit{direction changes} (major shifts in concept or design trajectory) than GPT sessions (1.6 vs.\ 0.4 per session), a ~4× increase, and more \textit{iterations} (distinct exploration branches) (3.3 vs.\ 2.5; Table~\ref{tab:metrics}). Anchoring was also weaker in \systemname{}: the L2 distance between first and last generated images in CLIP space was higher (M = 0.75, SD = 0.13) than under GPT (M = 0.46, SD = 0.31) as shown in Figure~\ref{fig:kreo_results_homogenization}, indicating endpoints less aligned with the starting image. Between-session spread in this distance was larger for GPT, consistent with more varied GPT use versus a more standardized \systemname{} pipeline.

Participants’ accounts help explain why. 
For example, several participants, including P9, P3, and P11, were satisfied with their initial image. In the GPT condition, early outputs were often treated as already complete. For example, P9 prompted a photorealistic kitchen and accepted the first result as ``presentable'' and ``pretty good,'' noting that the lack of alternatives limited further exploration. P12 similarly constructed a detailed prompt and described the returned image as ``something very finished right away,'' making only minor adjustments and treating it as ``close enough.'' In both cases, polished rendering appeared to signal completion. Rather than redirecting the image, participants tended to adjust prompts around the model’s first proposal.

In \systemname{}, participants described the opposite effect. Intermediate outputs were treated as provisional and open to change. After viewing several sketch-like outputs in the viewpoint stage, P1 described the images as “a question mark… something I can still go in and shape.” The rough, incomplete quality signaled that the image was still open to change, making it feel more accessible than a fully rendered result. P9 actively sketched on the canvas while developing a living room scene, adding a window and a corner detail that the system then interpreted and cleaned into a more precise drawing. He described these “sketchy images” as the “right level of detail to start... not too final”, using sketching was not just a way of editing the image but a way of thinking through the design.


These representations lowered the perceived cost of change, despite the cost of a model change being effectively the same, and made it easier to revisit earlier decisions without feeling that a finished solution had already been proposed.

\subsubsection{Staged interaction shifts users from evaluating outputs to constructing them}
Staged interaction shifted participants’ behavior from evaluating model outputs toward directly constructing and modifying them. \systemname{} sessions showed substantially more \textit{construct} actions (direct content specification or modification) than GPT (17.7\% vs.\ 4.3\%) indicating that participants more often directly modified the image rather than evaluating generated outputs, and a higher proportion of \textit{user-driven} actions (55.0\% vs.\ 25.8\%), indicating greater direct control over the artifact (Table~\ref{tab:metrics}). In contrast, GPT interaction remained more evaluation-heavy (39.4\% vs.\ 35.7\%).

Participants described this difference in terms of how they interacted with the image. P10, working on a small studio layout, alternated between erasing and redrawing parts of a counter and using masking to introduce a central island. Reflecting on this process, she noted that “drawing felt more direct than trying to phrase it.” Similarly, P13 used direct manipulation together with masking and prompting to rearrange the sofa in her living room, and said she preferred being able to move the couch “where I wanted it instead of describing it again.” 

This shift in interaction also changed how participants understood their role in the process. Rather than requesting outputs and reacting to them, participants described working on the image over time. P4, who in the GPT condition front-loaded detailed prompts to avoid iteration, contrasted this with \systemname{}, where he could modify individual elements without affecting the whole and being abele to ``finish 
what I've started.''
%
%
P6 described working with \systemname{} as “a conversation with a partner,” where each stage built on prior decisions. Across multiple stages, he developed a medieval-style kitchen from a black-and-white sketch, adjusted color and lighting, and then returned upstream to add a cat that propagated through the downstream layers. He expressed pride not just in the final image but in being able to recount how it came together: “I could say, first I did this, then I did that.” In contrast, although he described GPT outputs as ``gorgeous,'' he felt less ownership and said he ``would not be proud to show'' them because they felt too easy to produce.
Others expressed similar distinctions: P9 described the process as ``I worked with it, not just got something,'' while P2 noted that in GPT, ``I'm refining the prompt more than the image.''

 
Participants frequently framed their experience in terms of continuity and accumulation rather than repeated generation. As a result, authorship emerged from being able to trace and recount decisions rather than from the final image quality alone. This difference also shaped how participants valued the results. Although GPT outputs were often more polished or photorealistic, they were more likely to be treated as disposable. In contrast, participants expressed stronger attachment to \systemname{} outputs and greater interest in keeping or sharing them, suggesting that perceived ownership was tied to process contribution rather than output quality.


\subsubsection{Preserving decisions makes unintended changes visible and repairable}
While both systems exhibited similar levels of \textit{drift} (unintended deviations introduced by the model; 14.4\% in \systemname{}, 14.2\% in GPT), they differed in how these deviations were experienced and handled. While unintended drift occurred at similar rates in both conditions, \systemname{} showed slightly more intentional pivots (3.9\% vs.\ 2.9\%), suggesting that users more often redirected the process deliberately rather than accommodating model outputs. \systemname{} showed a higher \textit{revision burden} (ratio of \textit{repair} actions; 0.1 vs.\ $\sim$0) and non-zero \textit{invariant violations}, edits that unintentionally altered aspects outside their intended scope, (10.9\% vs.\ 0.0\%). . We note that these violations are participant-reported in the transcripts, rather than codings of observed behavior.

At first glance, this might suggest that \systemname{} introduces more errors. However, participant accounts indicate a different interpretation. 

For example, P15 prompted a country-style kitchen but received a more modern result and adopted it, explaining that the model had ``actually improved my taste... I shifted direction because of it.'' Similar reactions were reported by other participants, including P3, P6, and P13, who described baseline model deviations from their prompt as useful suggestions rather than errors.

This suggests that the increased revision burden in \systemname{} reflects not worse model performance, but a different interaction structure. By preserving prior decisions, the system makes inconsistencies explicit and localizable, allowing users to correct them directly. In contrast, one-shot generation obscures these inconsistencies by replacing the entire image at each step, and the high-resolution rendering style makes them feel like aesthetic improvements.

\subsubsection{Participants used staging as a flexible workspace rather than a fixed pipeline}

\begin{figure}[t]
\centering
\includegraphics[width=\linewidth]{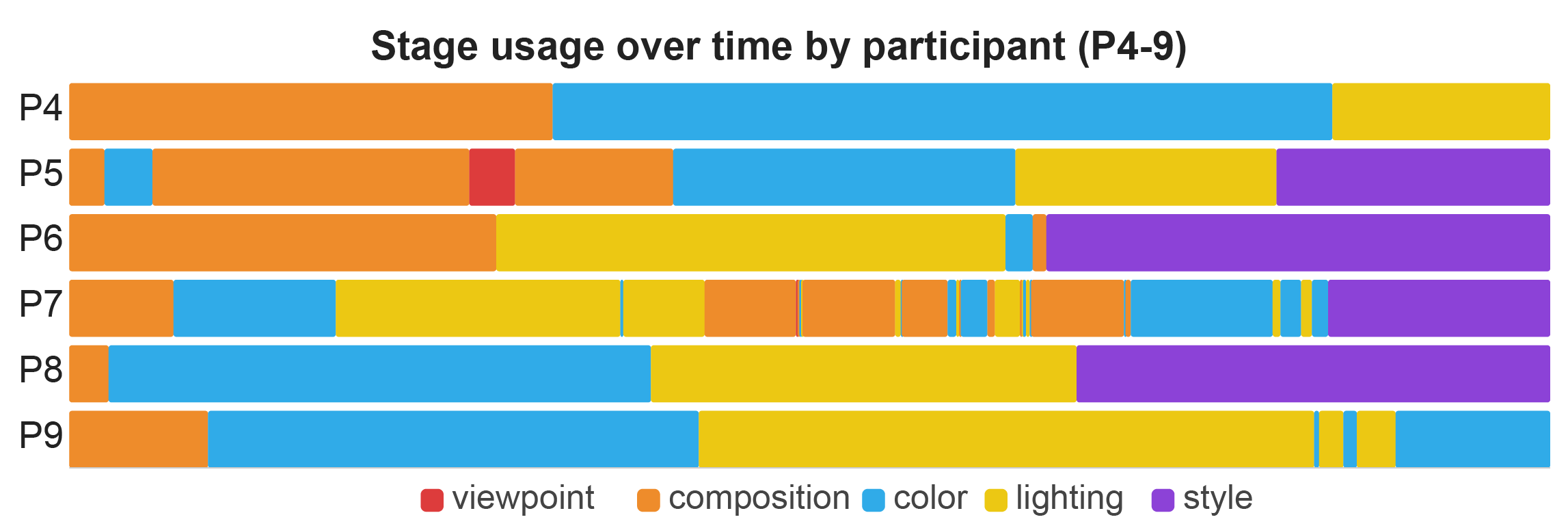}
\caption{
\systemname{} supports people's non-linear creative workflows. Participants P4-P9 frequently explored stages out of order and revisited stages to generate their images.
} 
\label{fig:kreo_usage_pattern}
\end{figure}

Participants did not follow the staged workflow linearly. A majority (66.7\%) revisited earlier stages and skipped others, entering through different starting points such as composition, lighting, or object-level refinement. These differences are reflected in stage usage patterns (Figure~\ref{fig:kreo_usage_pattern}). Interaction was most concentrated in Composition (39.9\% of actions), followed by Color (21.8\%), Lighting (19.5\%), and Style (18.8\%).

Participants also entered the workflow through different stages. Some began by focusing on composition or layout, while others prioritized lighting, color, or object-level refinement. For example, participants with backgrounds in architecture or 3D design often adjusted lighting early in the process, while others focused first on spatial arrangement or color relationships. These differences are reflected in stage usage patterns (Figure~\ref{fig:kreo_usage_pattern}), where interaction concentrated differently across stages depending on the participant.

Importantly, this structure appeared to align with participants’ existing mental models of image creation rather than imposing a new workflow. In the GPT condition, several participants independently decomposed their prompts into components such as viewpoint, composition, color, and style, effectively reconstructing a similar breakdown in natural language. Others externalized this process by using reference images or sketches before prompting, suggesting that they already conceptualized image creation as a sequence of separable decisions.

Together, these observations indicate that staging did not constrain users to a fixed pipeline, but instead provided a structured space that supported multiple valid entry points and trajectories. By making different dimensions of the image explicit and independently editable, \systemname{} enabled users to work in ways that reflected their prior experience, goals, and domain knowledge.

\section{Discussion and Future Work}
Our findings suggest that the main contribution of staged image generation is not simply increased controllability. It is a different interaction model. In the one-shot baseline, participants were more often positioned as evaluating and accommodating model proposals. In \systemname{}, they were more often able to build images through a sequence of revisable decisions whose effects remained visible over time. This suggests that an underexplored design dimension in generative systems is the temporal structure of interaction: not only what users can specify, but how a system supports intent formation, revision, and commitment over the course of creation.

\subsection{Progressive Commitment as a Temporal Interaction Strategy}
We interpret these results through the lens of \emph{progressive commitment}, in which artifacts evolve from coarse, under-specified representations to more detailed ones over time. Prior work in design and sketching shows that early, low-fidelity representations support exploration by keeping alternatives open and reducing the cost of change~\cite{tversky2002sketches,suwa2000unexpected,buxton2010sketching}. Our findings suggest that this principle carries over to generative systems. When users interacted with sketch-like intermediate outputs, they were more willing to redirect, revise, and elaborate their ideas. When they interacted with polished one-shot outputs, they were more likely to treat those outputs as proposals to assess or lightly tweak rather than as materials to reshape.

In this view, the contribution of \systemname{} is not simply that it decomposes image generation into several steps. It is that each step is paired with a representation suited to the decision at hand. Early stages support structural reasoning through provisional sketches, while later stages introduce appearance-level decisions such as color, lighting and style. This pairing helps users revise earlier choices without restarting the entire process and suggests that effective generative interfaces may depend as much on representational timing as on the number of controls they expose. 

At the same time, staging introduces new expectations. When visual dimensions are exposed as separable, users expect them to behave independently. Some of the tensions observed in the user study occurred when this expectation was only partially met, such as when a change intended for one stage affected another aspect of the image.  This highlights an important design challenge for future generative systems. 

Progressive interfaces do not only require more structure at the interaction level. They also require behavior that is sufficiently stable and localized to respect that structure. 

\subsection{Authorship and the Visibility of Creative Contribution}

Our findings also suggest that perceived authorship in generative systems depends on whether users can recognize their contributions, and chains of decisions in the resulting artifact. In one-shot workflows, participants often described themselves as selecting or evaluating outputs. In contrast, staged interaction made the construction process visible: users could trace how the image evolved through their actions, and describe a sequence of decisions leading to the final result.

This aligns with prior work showing that involving users in the creative loop increases perceived ownership~\cite{lubart2005can,guzdial2019interaction,shneiderman2007creativity,deterding2017mixed,oh2018lead}, but suggests that \emph{visibility of contribution} may be as important as participation itself. Systems that preserve intermediate decisions and make them inspectable may better support a sense of authorship, even when model assistance remains substantial.

An interesting implication of our findings is that \emph{these effects persist even when the cost of regeneration is the same for low-fidelity and high-fidelity outputs}. Prior work shows that people are less likely to critique high-fidelity representations and more willing to revise low-fidelity ones due to their perceived provisionality \cite{buxton2010sketching, goel1995sketches, tversky2002sketches, lim2008anatomy}. Our results suggest that this dynamic carries over to generative systems: even though generating a new image requires minimal effort, participants were less likely to redirect high-resolution outputs and more likely to explore when working with sketch-like representations. This extends prior findings from traditional design settings to generative AI, where the cost of producing alternatives is negligible but perceptions of finality continue to shape behavior.

\subsection{Limitations and Future Work}

Our prototype reflects one possible decomposition of visual decisions (e.g., viewpoint, composition, color, lighting, style), rather than a universal structure. Different domains may require alternative intermediate representations, and future work should explore how staging can be adapted to domain-specific workflows.

In addition, current generative models do not fully support clean separation between visual dimensions. Participants observed unintended interactions across stages, for example changing the color of the sofa could affect the shape, reflecting underlying entanglement in model representations. As staged interfaces make these dimensions explicit, improving controllability and disentanglement becomes increasingly important.

More broadly, progressive interfaces may extend beyond image generation to domains such as writing, programming, or decision-making, where users benefit from structuring complex tasks into revisable stages rather than producing final artifacts in one step. Additionally, our results suggest that interaction design—how outputs are staged, and how users can act on them—may be as important as model capability in shaping creative outcomes.

Finally, our study focuses on short ideation sessions. Many creative workflows unfold over longer timeframes or involve collaboration, where staged generation may support iterative refinement across sessions or coordination between contributors. Exploring progressive interaction in these settings remains an important direction for future work.

Overall, progressive commitment provides a promising framework for aligning generative systems with how ideas develop over time, supporting more exploratory, controllable, and personally meaningful forms of human–AI collaboration.

\section{Conclusion}
Current text-to-image systems generate fully detailed images in a single step, where many visual decisions are made together. As a result, editing one aspect often unintentionally changes others, and early outputs can anchor users before their ideas are fully formed.

We introduced \systemname{}, a multi-stage image ideation system that, inspired by how creative workflows isolate visual aspects, structures creation as a sequence of stages while preserving prior decisions. Instead of producing a final image upfront, \systemname{} begins with rough sketches and progressively adds detail. At each stage, users refine specific aspects of the image, while a locking mechanism ensures that earlier decisions remain stable during later edits.

In a comparative study, participants were better able to make targeted changes, explore alternatives, and trace how their decisions shaped the final image. This led to a stronger sense of ownership and produced less homogeneous outputs than one-shot generation.

These findings suggest that improving generative systems is not only about increasing control, but about structuring how and when decisions are made. Staged generation with mechanisms for preserving decisions can better support exploration, user agency, and diverse creative outcomes.

More broadly, our findings suggest that the design of generative systems should focus not only on model capability, but on how interaction is structured over time. By organizing generation around intermediate, revisable representations, progressive interfaces can better support how users form, refine, and take ownership of ideas. We see progressive commitment as a promising direction for building more exploratory, controllable, and human-centered generative tools.

\bibliographystyle{ACM-Reference-Format}
\bibliography{citations}

\clearpage
\newpage
\section{Appendix}
\appendix
\label{sec:appendix}

\section{\systemname{} Workflows: Two entry points to the same abstractions}
\begin{figure*}[t]
\centering
\includegraphics[width=\linewidth]{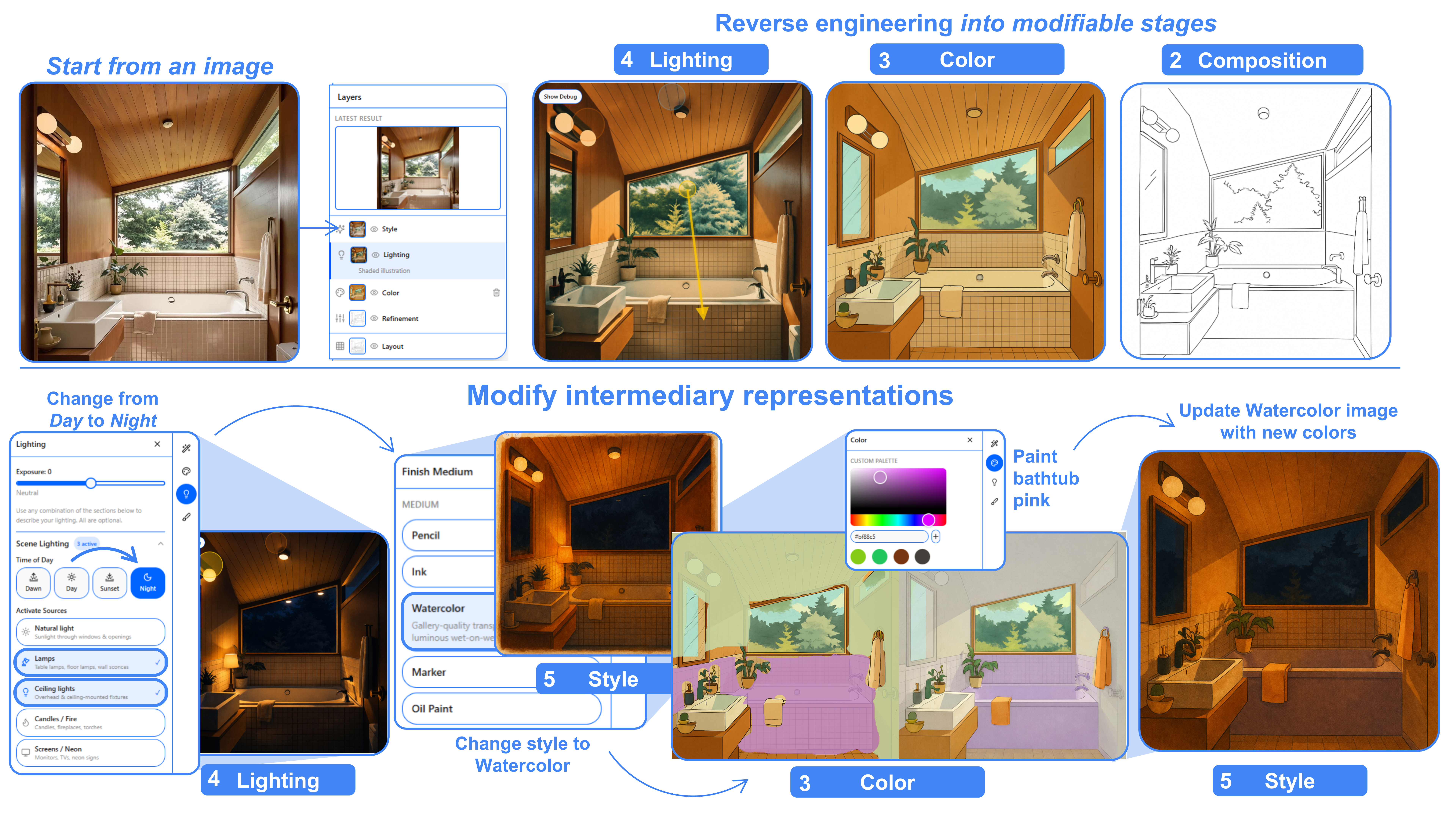}
\caption{Two entry points into \systemname{}'s progressive ideation workflow. Users can begin with 
an existing image. This image is reverse engineered into the \stage{lighting}, \stage{color} and \stage{composition} stages.
Users can edit each of the representations, as shown in the bottom half of the image, moving across stages and propagating decisions across stages. This allows for iterative refinement of visual decisions on existing images.}
\label{fig:kreo_entrypoints}
\end{figure*}

We support two entry points into the \systemname{} workflow: starting from a text prompt or from an existing image (Figure \ref{fig:kreo_entrypoints}).
In both cases, \systemname{} uses the same multi-stage generation process, allowing users to progressively manipulate independent representations.
However, in the prompt-first workflow, users begin with a natural-language description and start with a sketch.
Whereas, in the reference-first workflow, users upload an existing image and \systemname{} reverse-engineers it into the five stages.
This maps the finished image to the same intermediate representations used in the prompt-first workflow, allowing users to edit individual stages without regenerating the entire image.

\section{Example Workflow: Comic Illustration}

We walk through a case study to contextualize the design decisions we made in \systemname{} within real creative workflows.
In particular, we work with a comic illustrator, exploring how \systemname{} supports their process of designing a webcomic (\cref{fig:comics_figure}).

The illustrator's goal is to generate a cover image of their comic character, so they instantiate \systemname{} using a prompt-first workflow asking it to generate ``\textit{a close up of my main character}''.
While this is similar to traditional text-to-image systems, instead of generating a single high-fidelity image, \systemname{} starts by producing a set of six \stage{viewpoints} displaing the main character from above, below, and profile (\cref{fig:comics_figure}1).
These multiple versions scaffold the illustrators divergent thinking, allowing them to exploring different ways of viewing the scene without being influenced by stylistic details.

After selecting a viewpoint the matches their vision, the illustrator edits the sketch's \stage{composition}.
Since \systemname{} intentionally keeps the image as a simple sketch, the illustrator is empowered to draw, erase, move, and transform elements directly or apply masked AI-assisted edits to modify specific regions.
The starting viewpoint of the sketch not fully match their vision of the character, so the illustrator adjusts the proportions of the facial features,
—modifying the size of the eyes, reshaping the jawline, and refining the mouth—as well as parts of the back to better align with the intended viewpoint 
(\cref{fig:comics_figure}-Stage 2).
Since the image is represented as a sketch, they are able to quickly iterate on the character's facial features without the overhead of managing shadows or skin texture.

With the character's features now matching their expectations, the illustrator begins to increase image fidelity by adding \stage{color}.
In the color stage, users can define color palettes manually or generate them from prompts.
Since the illustrator draws the same character frequently, they directly select the colors they use for this characters, creating a palette of muted greens and blues (\cref{fig:comics_figure}-Stage 3).
They use direct manipulation tools to manually color in the character's hair with the highlights and lowlights they expect.
They use AI-assisted tools to fill in colors they are less particular about, converting a rough green stroke over the phone into a neatly colored phone.

Next, the illustrator decides to add shading and illumination in the \stage{lighting} stage.
This stage's representation now reflects depth, making it possible to see how lighting affects the scene.
The illustrator explores different light source placements, directions, and intensity, and explores different lighting conditions such as time of day or mood, finally settling on backlighting the character with neon lights (\cref{fig:comics_figure}-Stage 4).

Finally, the illustrator chooses to \stage{style} the image (\cref{fig:comics_figure}-Stage 5).
While \systemname{} supports various texture and stylistic effects (e.g., photorealistic, watercolor), the illustrator choosing digital painting to match their typical comic illustration style.
The composed image is now high-fidelity, but thanks to \systemname{}'s staged workflow, all of the illustrator's previous choices are maintained.
The character still reflects the facial structure they defined in composition and the hair details the illustrator sketched in color.

The final image only reflects the illustrator's design, not an AI model's assumptions.
Even though their original prompt did not specify decisions like the color of the character's hair or the neon background, \systemname{} did not fill that in for them.
Instead, the illustrator was able to make each of those decisions as they walked through each of \systemname{}'s stages, allowing them to leverage the power of text-to-image models while maintaining agency over their creative process.

\section{Example Workflow from the User Study}
We walk through a case study to contextualize the design decisions we made in \systemname{} within real creative workflows.

\begin{figure*}[t]
\centering
\includegraphics[width=\linewidth]{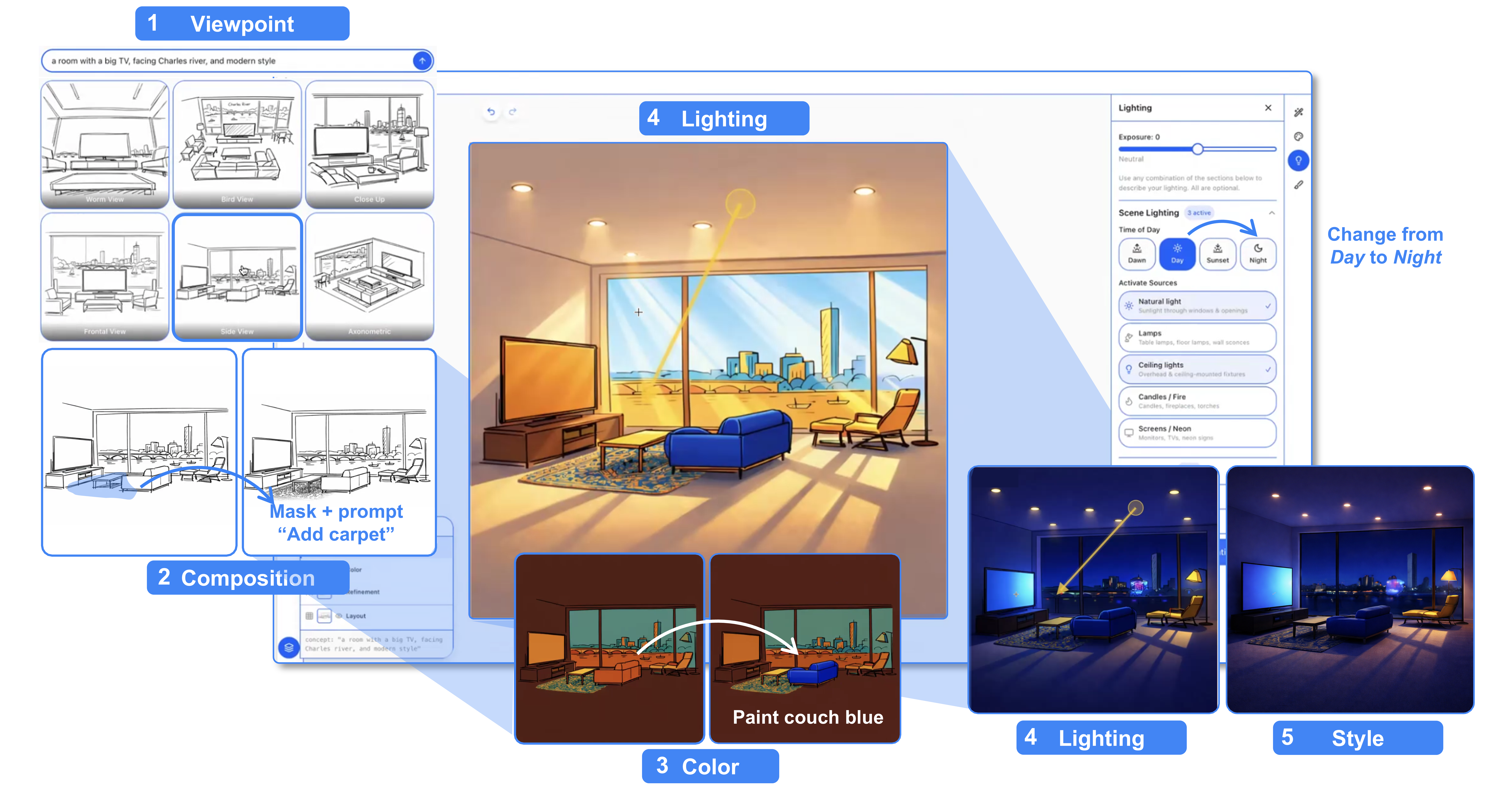}
\caption{
\systemname{} decomposes image generation into multiple stages. From a prompt, it generates (1) multiple \stage{viewpoints}, after which the illustrator (2) refines \stage{composition}, (3) \stage{color}, (4) \stage{lighting}, and (5) \stage{style} in any order. Stages can be completed in any order.}
\label{fig:interior_figure}
\end{figure*}

In particular, we work with an architecture student, exploring how \systemname{} supports their process of designing an interior living space (\cref{fig:interior_figure}).

The participant's goal is to generate a rendering of their ideal living room, so they instantiate \systemname{} using a structured prompt-first workflow, describing the scene in layers: the objects they want to see, the desired viewpoint, and the overall style.

While this is similar to traditional text-to-image systems, instead of generating a single high-fidelity image, \systemname{} starts by producing a set of \stage{viewpoint} sketches displaying the room from different angles and compositions (\cref{fig:interior_figure}-Stage 1).

These multiple versions scaffold the participant's divergent thinking, allowing them to explore different spatial arrangements without being influenced by stylistic or material details.

The participant gravitates toward a composition that preserves the Charles River view while keeping the interior visible, selecting it as the basis for further refinement.

With a viewpoint established, the participant begins editing the \stage{composition} of the scene.

Since \systemname{} intentionally keeps the image as a simple sketch, the participant is empowered to draw, erase, and apply masked AI-assisted edits to modify specific regions.

Using the masking tool, they identify an area of the room they want to enrich, and \systemname{} surfaces contextually relevant suggestions based on the region selected—recommending additions the participant had not explicitly considered, such as a carpet, but found immediately useful (\cref{fig:interior_figure}-Stage 2).

Since the image is represented as a sketch, they are able to quickly iterate on the spatial arrangement without the overhead of managing materials or lighting.

With the composition now matching their expectations, the participant begins to increase image fidelity by adding \stage{color}. In the color stage, users can define color palettes manually or generate them from prompts. The participant selects colors that reflect their personal aesthetic—applying a blue to the couch and introducing orange accents—using direct manipulation tools to paint specific regions of the scene (\cref{fig:interior_figure}-Stage 3). They use AI-assisted tools to reconcile the colors across the scene, noting that the model balances interior and exterior tones in a way that feels spatially coherent.
Next, the participant decides to add shading and illumination in the \stage{lighting} stage. This stage's representation now reflects depth, making it possible to see how lighting affects the interior space.

The participant explores different light source placements and directions, experimenting with multiple simultaneous sources and different times of day, and notes that this level of lighting control is something they had not previously encountered in generative tools (\cref{fig:interior_figure}-Stage 4). They observe that \systemname{} accurately propagates shadows consistent with the light directions they specify, reflecting an understanding of 3D spatial logic despite operating on a 2D image.

Finally, the participant chooses to \stage{style} the image (\cref{fig:interior_figure}-Stage 5). While \systemname{} supports various texture and stylistic effects, the participant selects a photorealistic finish to match the kind of architectural rendering they produce in their professional workflow. The composed image is now high-fidelity, but thanks to \systemname{}'s staged workflow, all of the participant's previous choices are maintained. The room still reflects the spatial composition they defined in the composition stage and the material palette they specified in color. 

Even though their original prompt did not specify decisions like the color of the couch or the direction of the light, \systemname{} did not fill those in for them. Instead, the participant was able to make each of those decisions as they walked through each of \systemname{}'s stages, allowing them to leverage the power of generative image models while maintaining agency over their creative process—something they noted was a meaningful departure from the prompt-and-wait workflows they typically rely on.

\section{User Study Protocol and Analysis Details}
\label{sec:study_appendix}

This appendix provides additional methodological details for the user study described in Section~\ref{sec:userstudy}, including participant recruitment, study procedure, materials, and analysis methods.







\subsection{Participant Recruitment and Screening}

Participants were recruited through posts in online communities focused on illustration, art, and design. Interested individuals completed a short pre-screening survey about their creative background, experience with generative image tools, and availability.

A total of 275 responses were collected. From these responses, we invited 19 participants whose profiles met our study criteria and represented a range of visual design domains including illustration and concept art, interior design and architectural visualization, comics and storyboarding, graphic design and marketing, and animation or motion design.

Participants were required to be at least 18 years old and have prior hands-on experience using generative image tools as part of a creative workflow. The screening survey also collected information about which tools participants used (e.g., Midjourney, Stable Diffusion, DALL·E, Adobe Firefly), how frequently they used them, and at which stages of their creative process they were typically employed. Participants received a gift card as compensation for their time.

\subsection{Study Design}

The study used a within-subjects design in which each participant completed two creative tasks using two different tools:

\begin{itemize}
\item \textbf{Progressive staging interface (\systemname{})}
\item \textbf{One-shot text-to-image baseline}
\end{itemize}

To prevent participants from refining the same concept across conditions, we used two related prompts:

\begin{itemize}
\item \textit{Design your ideal living room}
\item \textit{Design your ideal kitchen}
\end{itemize}

Each participant completed both prompts, using one prompt per tool. Tool order and prompt order were counterbalanced across participants using four sequences (Table~\ref{tab:counterbalance}).

\begin{table}[h]
\centering
\begin{tabular}{ccc}
\hline
Sequence & Tool Order & Prompt Order \\
\hline
A & \systemname{} $\rightarrow$ Baseline & Living Room $\rightarrow$ Kitchen \\
B & \systemname{} $\rightarrow$ Baseline & Kitchen $\rightarrow$ Living Room \\
C & Baseline $\rightarrow$ \systemname{} & Living Room $\rightarrow$ Kitchen \\
D & Baseline $\rightarrow$ \systemname{} & Kitchen $\rightarrow$ Living Room \\
\hline
\end{tabular}
\caption{Counterbalancing design used in the study.}
\label{tab:counterbalance}
\end{table}

\subsection{Session Structure}

Each session lasted approximately 60 minutes and followed the structure below:

\begin{itemize}
\item Consent and study overview (4 minutes)
\item Screener questions and workflow warm-up discussion (6 minutes)
\item Tool A introduction (2 minutes)
\item Tool A task (14 minutes)
\item Post-condition reflection (7 minutes)
\item Tool B introduction (2 minutes)
\item Tool B task (14 minutes)
\item Post-condition reflection (7 minutes)
\item Final comparison and wrap-up (4 minutes)
\end{itemize}

Sessions were conducted remotely via video conferencing with screen sharing enabled.

\subsection{Task Instructions}

Participants were asked to generate one image per condition using the assigned prompt. During the task, participants were instructed to think aloud and describe what they were noticing, attempting, or deciding between while working.

Moderators occasionally used light probes to clarify participants’ intentions, such as asking what they were trying to change, what they expected to happen, or why they shifted direction.

Examples of probes included:
\begin{itemize}
\item ``What are you trying right now?''
\item ``What made you try that?''
\item ``What were you hoping would change?''
\item ``Why did you decide to keep this result?''
\end{itemize}

After each condition, participants completed a short reflection discussing the outcome they produced, how the idea evolved during the task, and how the workflow compared to their usual creative process. After both conditions were completed, participants participated in a final comparison discussion about differences between the two systems.

\subsection{Materials and Setup}

The study used two tools:

\begin{itemize}
\item \systemname{}, accessed through a browser-based interface.
\item A prompt-centric one-shot text-to-image interface supporting iterative prompting and image generation.
\end{itemize}

Participants interacted with both tools via screen sharing. Screen and audio recordings were captured for later analysis. For each session we also collected interaction artifacts including generated images, prompt histories, and stage histories produced during the session.


We collected both interaction data and participant reflections.

\subsection{User Study Analysis Details}
\label{sec:study_appendix}

This section describes the data processing, annotation scheme, and metric definitions used in our mixed-methods analysis comparing progressive staging (\systemname{}) and one-shot workflows.

\textbf{Data Sources and Processing}

We analyzed screen recordings and audio from each session to reconstruct interaction histories, including prompts, edits, generated outputs, and participants’ verbal reasoning. Sessions were segmented into discrete \textit{actions} (e.g., prompting, modifying content, evaluating outputs, or correcting unintended changes).

\subsubsection{Research Questions and Metrics}

Our analysis is structured around three research questions:

\textbf{RQ1: Exploration vs. Anchoring.}
We examine how users explore and revise ideas during early ideation. Anchoring is measured via similarity between initial and final outputs and the number of direction changes. Exploration is characterized through iteration structure and the proportions of on-intent, pivot, and drift actions.

\textbf{RQ2: Control and Predictability.}
We evaluate how disentangling decisions affects control over outcomes. This includes unintended changes (invariant violations), user-driven interaction, and revision effort.

\textbf{RQ3: Non-linear Workflows.}
We analyze how users appropriate staged interaction, including stage skipping, revisiting earlier stages, and iterative refinement through repropagation.

\subsubsection{Action Annotation}

Each action was manually annotated along the following dimensions:

\textbf{Action type.}
\begin{itemize}
    \item \textbf{Construct:} Direct specification or modification of content
    \item \textbf{Evaluate:} Inspecting or assessing outputs
    \item \textbf{Generate:} Producing new outputs via prompting or regeneration
    \item \textbf{Refine:} Incremental adjustment of an existing idea
    \item \textbf{Repair:} Correcting unintended changes
\end{itemize}

\textbf{Intent.}
\begin{itemize}
    \item \textbf{On-intent:} Aligns with current design direction
    \item \textbf{Pivot:} Deliberate shift to a new direction
    \item \textbf{Drift:} Unintended deviation introduced by the model
\end{itemize}

\textbf{Agency.}
\begin{itemize}
    \item \textbf{User-driven:} Direct specification or modification
    \item \textbf{Model-led:} Generating and reacting to outputs
\end{itemize}

\textbf{Additional annotations.}
\begin{itemize}
    \item \textbf{Direction change:} Major shift in concept or trajectory
    \item \textbf{Invariant violation:} Unintended changes outside intended scope
    \item \textbf{Iteration ID:} Groups actions into exploration branches
\end{itemize}

\subsection{Metric Definitions}

All metrics are computed at the session level.

\textbf{Exploration and anchoring.}
\begin{itemize}
    \item \textbf{Direction changes:} Number of major shifts in trajectory
    \item \textbf{Exploration breadth:} Number of distinct iterations
    \item \textbf{Concept drift:} Proportion of \textit{drift} actions
    \item \textbf{Intentional pivots:} Proportion of \textit{pivot} actions
\end{itemize}

\textbf{Anchoring (image similarity).}
We compute cosine similarity between embeddings of the first and final generated images using OpenCLIP with L2-normalized embeddings.

\textbf{Control and predictability.}
\begin{itemize}
    \item \textbf{Constructive engagement:} Proportion of \textit{construct} actions
    \item \textbf{Evaluation-heavy behavior:} Proportion of \textit{evaluate} actions
    \item \textbf{Perceived agency:} Proportion of \textit{user-driven} actions
    \item \textbf{Revision burden:} Ratio of \textit{repair} actions
    \item \textbf{Invariant violations:} Proportion of violations
\end{itemize}

\textbf{Workflow structure.}
\begin{itemize}
 \setlength{\itemsep}{2pt}
    \setlength{\parskip}{0pt}
    \setlength{\parsep}{0pt}
    \item \textbf{Iteration structure:} Number and distribution of iterations
    \item \textbf{Stage transitions (\systemname{}):} Revisiting and skipping stages
    \item \textbf{Stage usage:} Distribution and adoption across stages
\end{itemize}

\textbf{Aggregation} Metrics are computed per session and aggregated across participants, reporting mean and standard deviation per condition.

\subsubsection{Qualitative Analysis}
We conducted a thematic analysis of think-aloud protocols and post-task interviews to understand experiences of authorship, control, predictability, and commitment. Themes were developed iteratively and used to interpret quantitative patterns.

\subsection{Scope and Limitations}
Because session durations were not consistently available, all metrics are based on action counts and proportions rather than time-based measures. Additionally, some constructs (e.g., perceived agency) are approximated using behavioral proxies derived from interaction logs.

\section{System Design}
\begin{figure}
\centering
\includegraphics[width=\linewidth]{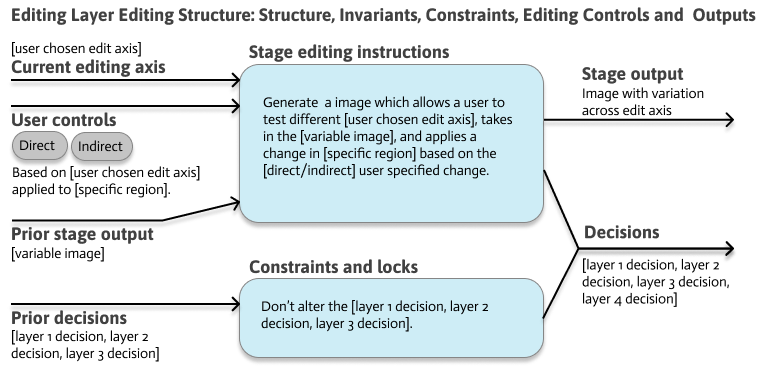}
\caption{Diagram depicting how editing instructions and constrains are dynamically passed to each layer.}
\label{fig:editing_contraints_propagation_diagram}
\end{figure}

\begin{figure*}
\centering
\includegraphics[width=\linewidth]{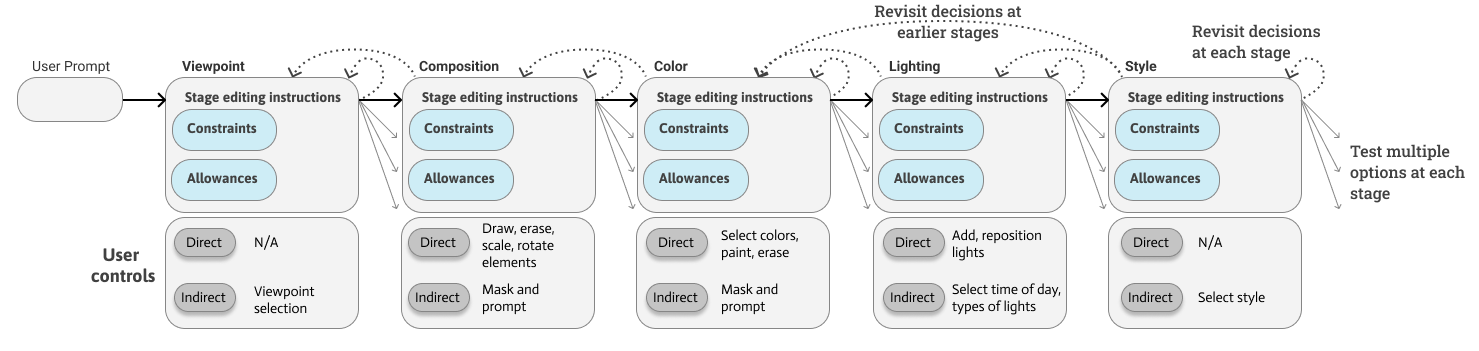}
\caption{Diagram depicting how editing instructions and constrains passed across layers; revision loops etc.}
\label{fig:decision_propagation_diagram}
\end{figure*}

The figures depict inputs, outputs, user controls and constraints at each stage. Figure \ref{fig:editing_contraints_propagation_diagram} depicts the system design of inputs, outputs and constraints at each stage, while Figure \ref{fig:decision_propagation_diagram} depicts how editing instructions are passed across each of the 5 layers.

\end{document}